\documentclass[journal,a4paper]{IEEEtran}
\IEEEoverridecommandlockouts
\usepackage{comment}
\usepackage{cite}
\usepackage{amsmath,amssymb,amsfonts}
\usepackage{algorithmic}
\usepackage{graphicx}
\usepackage{textcomp}
\usepackage{xcolor}
\usepackage{url}
\usepackage{multirow}
\usepackage{subfigure}
\usepackage[normalem]{ulem}
\usepackage{soul}

\newcommand{\R}{\mathrm{R}}
\newcommand{\T}{\mathrm{T}}

\renewcommand{\d}{\mathrm{d}}
\renewcommand{\r}{\mathrm{r}}

\begin{document}

\setlength{\oddsidemargin}{-0.45in}
\setlength{\textwidth}{43pc}
\setlength{\textheight}{58pc}
\setlength{\topmargin}{-49.0pt}

\title{THz Channels for Short-Range Mobile Networks: Multipath Channel Behavior and Human Body Shadowing Effects}

\author{Minseok~Kim, Jun-ichi~Takada, Minghe~Mao, Che~Chia~Kang, Xin~Du, and Anirban~Ghosh 
\thanks{Minseok Kim and Minghe Mao are with the Graduate School of Science and Technology, Niigata University, Niigata 950-2181, Japan (e-mail: mskim@eng.niigata-u.ac.jp).}
\thanks{Jun-ichi Takada is with the Office of Global Affairs, Institute of Science Tokyo, Tokyo 152-8550, Japan.}
\thanks{Che Chia Kang and Xin Du were with the Department of Transdisciplinary Science and Engineering, Tokyo Institute of
Technology (formerly Institute of Science Tokyo), Tokyo 152-8550, Japan.}
\thanks{Xin Du is currently with the Graduate School of Science and Engineering, Kagoshima University, Kagoshima 890-0065, Japan.}
\thanks{Anirban Ghosh is with the Department of Electronics and Communication Engineering, SRM University AP, Andhra Pradesh 522240, India.}
\thanks{This work was supported by the Commissioned Research through the National Institute of Information and Communications Technology (NICT) (\#JPJ012368C02701), and the Ministry of Internal Affairs and Communications (MIC)/FORWARD (\#JPMI240410003), Japan.}
}

\maketitle

\begin{abstract}
The THz band (0.1--10 THz) is emerging as a crucial enabler for sixth-generation (6G) mobile communication systems, overcoming the limitations of current technologies and unlocking new opportunities for low-latency and ultra-high-speed communications by utilizing several tens of GHz transmission bandwidths. However, extremely high spreading losses and various interaction losses pose significant challenges to establishing reliable communication coverage, while human body shadowing further complicates maintaining stable communication links. Although point-to-point (P2P) fixed wireless access in the THz band has been successfully demonstrated, realizing fully mobile and reliable wireless access via highly directional communication remains a challenge. This paper addresses the key challenges faced by THz mobile networks, focusing particularly on the behavior of multipath channels and the impact of human body shadowing (HBS). It presents the environment-dependent characteristics of multipath clusters through empirical measurements at 300~GHz using a consistent setup, highlighting the need to account for environmental factors in system design. In addition, it presents a motion capture-based approach for precise measurement and prediction of HBS to enable proactive path scheduling and enhances link reliability, offering key insights for robust THz communication systems in future 6G networks.
\end{abstract}



\section{Introduction}
While the deployment of fifth-generation (5G) mobile networks at Frequency Range 1 (FR1) (sub-6 GHz) and FR2 (24.25--52.6 GHz) is accelerating globally, the research and development focus is already shifting towards beyond 5G (B5G) and sixth-generation (6G) mobile networks. Future mobile networks are expected to demand low-latency of less than 1~ms and ultra-high data rates exceeding 100 Gbps, more than ten times faster than 5G. Recently, utilizing FR3 (mid-band, 7.1--24.25 GHz) has gained attention, but developing terahertz (THz) bands above 100 GHz (100 GHz--10 THz), which still can leverage large channel bandwidths of several tens of GHz, remains crucial. As a result of the World Radiocommunication Conference 2019 (WRC-19), a part of the THz spectrum, 137 GHz within the 275--450 GHz band, has been identified for land mobile and fixed services, enabling ultra-high-speed communications at the 100 Gbps level. Further, at the WRC-23 in Dubai, a proposal to allocate the frequency bands of 102--109.5 GHz, 151.5--164 GHz, 167--174.8 GHz, 209--226 GHz, and 252--275 GHz for the future development of International Mobile Telecommunications (IMT) was approved as a preliminary agenda item for WRC-31. This increases expectations for the utilization of low-THz or sub-THz bands (100--300~GHz) beyond 2030. To effectively use THz waves in mobile networks, it is crucial to develop highly directional transmission methods that can overcome significant spreading losses. Until now, THz waves have primarily been explored for point-to-point (P2P) fixed and backhaul communications. However, a document published in 2024 by the THz ETSI Industry Specification Group (ISG) identified 19 use cases in which THz communications are expected to play a major role, with short-range mobile access also being recognized as a promising application. 

\begin{figure}[t]
\centering
\includegraphics[width=0.9\linewidth]{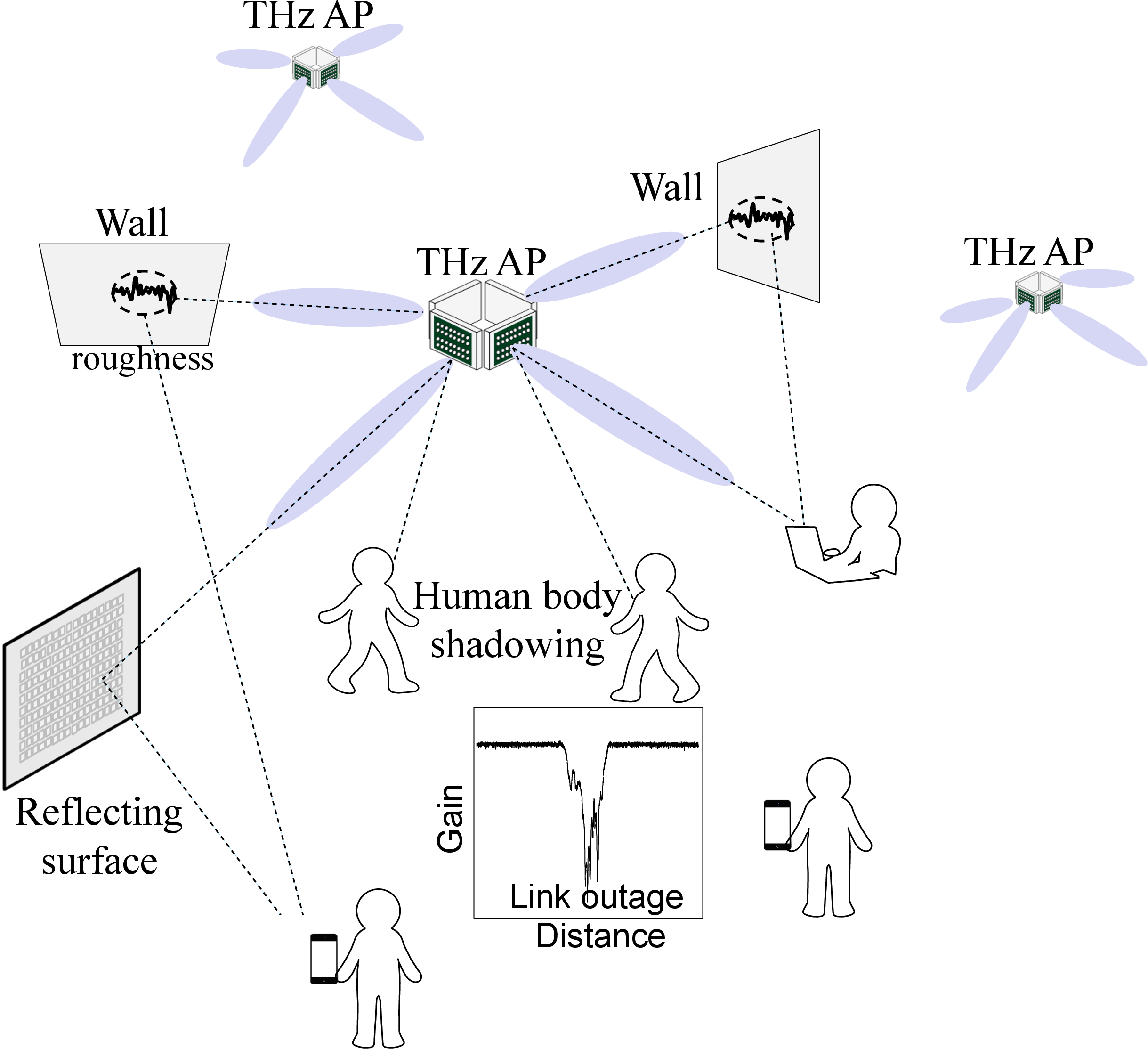} 
\caption{A potential application of THz mobile networks. \label{fig:concept}}
\end{figure}

Fig.~\ref{fig:concept} shows a potential application of THz mobile networks. The pronounced line-of-sight (LoS) propagation and significantly large propagation losses in THz waves make it difficult to establish reliable communication coverage, especially in non-line-of-sight (NLoS) regions, due to shadowing caused by both static and dynamic objects. To achieve fully mobile and reliable wireless access, highly directional communication that leverages multipath propagation is essential. However, significant reflection losses, including the impact of rough surface scattering, greatly reduce the chances of obtaining viable multipath signals as the distance between antennas and reflectors increases. Consequently, securing multipaths suitable for multi-stream data transmission and diversity combining becomes challenging. Expanding coverage by densely deploying access points or base stations is a straightforward solution, but actively leveraging multipath propagation is essential to reduce installation and maintenance costs. Thus, a transmission approach that dynamically manages multiple beams along both direct and reflected paths to increase channel capacity and maintain communication continuity despite human body blockage is highly desirable. Site-specific deployment of passive or intelligent reflecting surfaces (PRS/IRS) is crucial for maximizing multipath utilization. While IRS holds promise for improving propagation, challenges remain in achieving real-time wave control, optimal placement in complex environments, and developing reliable, cost-effective hardware for THz operation. Accordingly, as a prerequisite to the deployment of such advanced technologies, it is essential to investigate the intrinsic properties of THz multipath channels. 

This paper investigates the key challenges associated with the deployment of THz mobile networks, focusing on the behavior of multipath channels and the effects of HBS. Environment-dependent characteristics of multipath clusters are examined through measurements at 300~GHz. In contrast to existing studies that focused on a specific environment, thereby limiting the generalizability of their findings, this work systematically analyzes channel behavior across several typical environments using a consistent measurement setup. The results highlight the importance of incorporating environmental factors into system design, particularly for enabling spatial multiplexing and diversity techniques. Furthermore, the paper presents a motion capture-based method for precise measurement and prediction of HBS that incorporate human-shaped screen modeling. This technique enables accurate blockage detection and facilitates proactive scheduling of propagation paths to mitigate shadowing-induced link degradation. The presented results provide valuable insights for the design of robust, high-data-rate, short-range THz communication systems anticipated in future 6G wireless networks.

The remainder of this article is organized as follows. Section II provides a brief review of recent progress in the field. Section III presents multipath cluster channel characteristics obtained from double-directional high-resolution channel sounding at 300~GHz across a variety of indoor and outdoor mobile access scenarios, covering distances of approximately 5 to 50 meters. These results are crucial for estimating the potential performance of THz propagation in practical deployment environments. Section IV proposes a ray-tracing-based HBS model using a human-shaped screen from motion capture data for accurate THz link prediction. Section V discusses the future research direction, and Section VI concludes the paper.

\section{Recent Progress}
The modeling of the disparate properties of the THz band compared to the other well-studied lower-frequency spectrum may need a fresh approach. In this context, accurate channel modeling is essential for the reproducibility of radio wave propagation, validation of novel algorithms, and development of new systems. In this section, recent progress in channel measurement and modeling of multipath cluster channels at THz frequencies and HBS at THz and millimeter-wave (mm-wave) frequencies is overviewed. 

\subsection{Multipath Cluster Channels}
Channel measurements followed by modeling and characterization using various approaches have been reported in the literature for the THz band. In indoor settings, the environment that has been in focus includes data centers \cite{datacenter}, offices \cite{office}, conference rooms \cite{conference}, and corridors \cite{TAP}. In \cite{datacenter}, a modified Saleh-Valenzuela (S-V) model, where the modification lies in defining the cluster power as a function of delay, is used for channel modeling. In comparison, \cite{office} uses an extended S-V model, which is dependent on the frequency of operation and stochastic properties of the position of transmitter (Tx) and receiver (Rx). 

Channel modeling can be broadly classified into stochastic, deterministic, and hybrid models. Hybrid channel modeling combines either the strengths of both deterministic and stochastic approaches or different deterministic approaches to balance accuracy and computational efficiency. In the quasi-deterministic (QD) model, a widely accepted hybrid modeling approach, the dominant multipath components (MPCs) are deterministically derived from the environmental geometry, whereas scattered MPCs from occasional small and random objects are stochastically represented as random clusters. For each cluster, dispersion in the delay and angular domains is subsequently incorporated using an intra-cluster stochastic model. This approach has been successfully applied to THz propagation in various indoor and outdoor scenarios, providing an optimal trade-off between accuracy and complexity \cite{conference, TAP}.

In non-indoor settings such as train to infrastructure (T2I) \cite{rail} and urban microcell (UMi) \cite{urban}, multipath clustering has been used for modeling signal propagation in the THz band. While a third-generation partnership project (3GPP)-like QD model is used in \cite{rail}, \cite{urban} presents a spatial autocorrelation function that can be used to develop a spatially consistent channel model. Individual and concerted efforts across the globe have led to measurement campaigns in the low THz band in different environments. Extensive measurement campaigns for parameterization remain insufficient, and a standardized channel model capable of accurately capturing cluster channel characteristics in the THz bands across diverse application scenarios has yet to be established.

\subsection{Human Body Shadowing}
To the best of the authors' knowledge, limited literature is currently available on HBS measurements in the THz band. In \cite{Peize}, human shadowing measurements were compared with the Double Knife-Edge Diffraction (DKED) and METIS models. Although discrepancies in the predicted shadowing duration were observed, the study did not comprehensively investigate potential sources of error, such as positioning inaccuracies or limitations in the propagation models, thereby limiting the ability to fully assess the reliability of the predictions. The study in \cite{Hirata} proposed complex screen models based on the Knife-Edge Diffraction (KED) theorem to represent a specific human phantom, and demonstrated good predictive accuracy. However, the models were specialized for the particular phantom used, limiting their applicability to general human-induced blockages. These observations collectively highlight the need for a more generalized and physically grounded modeling framework capable of accurately characterizing arbitrary human shadowing scenarios.

On the other hand, several studies have investigated HBS effects in the mm-wave bands. In particular, the measurements for the propagation channel affected by a pedestrian have been conducted \cite{NISTHBS, SanaHBSmulti} using continuous wave (CW) signals. HBS properties, including shadowing loss and shadowing duration, were empirically parameterized from measurements of a fixed propagation channel shadowed by a human body moving at a constant speed. Given the short wavelength of mm-wave and THz bands, these HBS properties can vary significantly with instantaneous human motion. Therefore, further investigation and measurement of the relationship between the dynamic channel and corresponding human motion are required.

Various human models for deterministic simulations have been proposed and reviewed in \cite{NISTHBS}. In these simulations, human blockers can be represented by rectangular screens, spheres, cylinders, and polygons. However, in the THz band, ray tracing (RT)-based simulations often suffer from limited accuracy, while electromagnetics (EM)-based simulations incur prohibitively high computational costs. Consequently, a balanced approach that incorporates a detailed human body model with RT-based simulation is essential.

\section{Multipath Cluster Channel Measurement and Modeling}
Integrating multiple-input-multiple-output (MIMO) antennas into THz systems enables ultra-narrow directional beamforming to mitigate severe path loss, and multi-stream transmission via multi-beam MIMO can significantly enhance data rates. Consequently, understanding multipath cluster channel characteristics at THz frequencies is essential for accurately assessing the intrinsic channel capacity of an environment and expanding coverage with PRS/IRS. This section introduces an in-house developed channel sounder and presents the environment-dependent multipath cluster channel properties measured at 300 GHz across various indoor and outdoor scenarios.

\subsection{Double-Directional Channel Sounder}
Channel sounding captures samples of the transfer function in time, frequency, and space domains to characterize the dispersive properties of a multipath propagation channel in terms of Doppler frequency, delay, and angles. These sampling domains are interchangeable due to the Fourier transform duality relationships, linking Doppler frequency with time, delay time with frequency, and angle with space.

Regarding angle domain channel acquisition, there are three widely recognized approaches: a full array, a virtual array, and angle scanning with directive antennas. In a full-array setup, simultaneous transmission from multiple antennas using orthogonal waveforms enables fast acquisition in the space domain. However, this approach typically involves greater complexity and higher development costs. On the other hand, a virtual array can capture channel responses in the space domain using a single antenna element, which is physically moved to various positions to form a synthetic array. While this method requires more time for measurement, it is generally more cost-effective. Both of these approaches rely on array signal processing, such as beamforming, to obtain channel responses in the angle domain, which is equivalent to performing a Fourier transform on the samples in the space domain. Angle scanning, a preferred method for channel measurements for mm-wave and THz bands, can relax post-processing requirements. This is beneficial because virtual arrays often experience phase drift over long measurement periods due to the relatively high phase noise of local oscillators. However, prolonged measurement time limits the ability to capture Doppler characteristics in dynamic scenarios. Thus, both the virtual array and angle scanning methods should maintain a static channel condition during measurement. 

Several studies have developed channel sounders for THz bands. In \cite{Access2}, an instrument-based development was proposed. On the Tx side, an intermediate frequency (IF) signal with an 8 GHz bandwidth is generated by an arbitrary waveform generator and upconverted to 300 GHz. On the Rx side, the downconverted IF signals are sampled using an ultra-high-speed digitizer with a rate of 32 GSa/s. Angular characteristics are obtained by panning and tilting 26 dBi rectangular horn antennas on both the Tx and Rx sides. Due to the limited output power, the equivalent isotropically radiated power (EIRP) is around 16 dBm, which restricts the measurable distance. In a static environment, however, the signal-to-noise ratio (SNR) can be enhanced by averaging multiple sounding symbols, resulting in a gain of $10\log_{10}M$~dB, where $M$ is the number of averages. A dynamic range of approximately 60 dB is achieved when 100 snapshots are averaged, with a minimum separation distance of 1 meter; thus, multipaths can be measured at Tx-Rx separation extending to several hundred meters.

\begin{figure*}[t]
\centering
\subfigure[Omnidirectional power delay profiles (PDPs).\label{fig:PDPs}]{\includegraphics[width=0.61\linewidth]{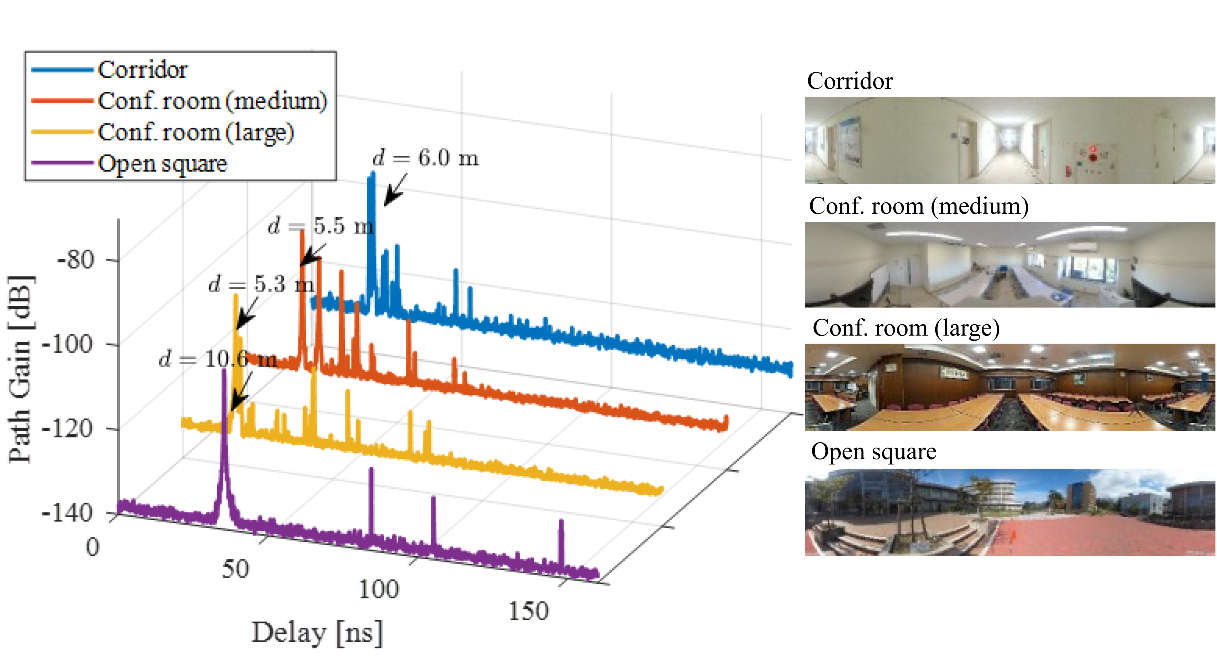}} 
\subfigure[Number (upper) and relative power (lower) of clusters.\label{fig:CDF_CL}]{\includegraphics[width=0.38\linewidth]{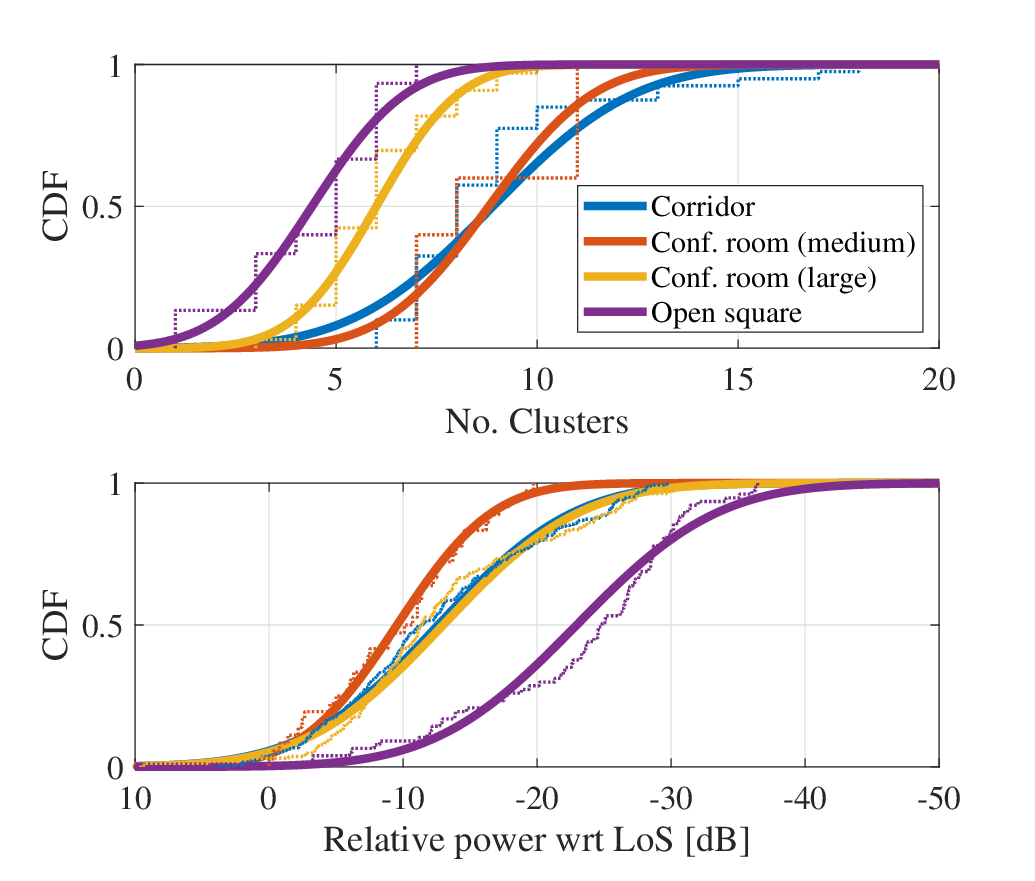}}
\caption{Environment dependent cluster characteristics.\label{fig:Measurement}}
\end{figure*}

\subsection{MPC vs. Multipath Cluster}
MPCs refer to individual propagation paths, while multipath clusters represent groups of MPCs with similar delays and angles, typically corresponding to unresolvable MPCs from nearby scatterers. The S–V model is a classic stochastic channel model that characterizes MPCs based on clustering in the delay domain. More recently, Geometry-based Stochastic Channel Models (GSCMs) including 3GPP TR~38.901 have extended this concept by incorporating both delay and angular characteristics. These models have become the standard framework for accurately capturing realistic propagation behavior in modern mobile network design. The concept of multipath clusters provides a more manageable and realistic representation of the channel structure that is both spatiotemporally consistent and physically meaningful. 

However, measurement results in the THz bands indicate that several specular MPCs are sparsely distributed and clearly separable in both the delay and angular domains \cite{Access2, TAP, WCL}, suggesting that it is no longer necessary to treat MPCs as clusters in such cases. This is primarily because propagation in the THz band is dominated by specular reflections, even from small objects, due to the extremely short wavelengths. Furthermore, highly directional THz communication, combined with wider bandwidth, enables significantly higher spatiotemporal resolution, allowing individual MPCs to be more easily distinguished. Nevertheless, diffuse scattering caused by roughness of reflecting surfaces can still contribute significantly to channel fading, and can be viewed as a cluster of unresolvable MPCs, whose delays and angles are closely aligned with those of the associated specular reflection. Cluster channel modeling and its validation in the THz band remain crucial research directions, with microscopic interactions generating intra-cluster behavior.

\subsection{Environment-Dependent Cluster Channel Characteristics}
To investigate the environment-dependent trends, both indoor scenarios, such as two different-scale conference rooms \cite{Access2} and a corridor \cite{TAP}, and an outdoor scenario, an open square hotspot \cite{WCL}, were selected as typical environments, as shown in Fig.~\ref{fig:PDPs}. The analysis was conducted based on cluster extraction and power spectrum synthesis, utilizing double-directional channel impulse responses obtained through angle scanning measurements. The key findings from the analysis of the 300 GHz band measurement data are summarized as follows. 

The primary propagation mechanism observed, aside from the direct wave, is single-bounce reflection.  The power delay spread characteristics obtained through omnidirectional antenna pattern synthesis are highly dependent on the environment and generally less than $10$~ns, as shown in Fig.~\ref{fig:PDPs}. The maximum excess delay is less than $100$~ns in indoor environments and $160$~ns in the open square environment. As the scale of the environment increases, the distance between the antenna and scatterers grows, causing a decrease in the number of multiple reflected waves with substantial power.

The upper figure of Fig.~\ref{fig:CDF_CL} illustrates the cumulative distribution of the number of clusters, which increases from left to right in the order of open square, large conference room, medium conference room, and corridor. In terms of average values, corridors, and medium conference rooms have eight clusters, large conference rooms have six, and open squares have four, indicating a tendency for the number of clusters to decrease as the environment size increases. Regarding the power of multipath clusters, the lower figure of Fig.~\ref{fig:CDF_CL} shows the cumulative distribution of relative power compared to the LoS path. That increases in the order of medium conference room, corridor, large conference room, and open square. The proportion of multipath clusters with a relative power greater than $-10$ dB, which could be advantageous for multi-stream transmission, is approximately 40~\% in indoor environments. However, in open square environments, this proportion drops below 10~\%, making multipath utilization less feasible in such outdoor environments.

Propagation loss characteristics improved by several dB compared to free-space path loss (FSPL) due to the presence of multipaths. The distance between the antenna and the scatterer becomes larger, so the number of observable multipath clusters becomes even smaller. In particular, in the open square environment \cite{WCL}, single-bounce reflected waves from glass walls and windows were prominent. In NLoS conditions, diffracted waves other than reflected waves cannot be observed, making it difficult to receive signals. An obstruction by tree leaves caused a loss of larger than $10$ dB.

\subsection{Cluster Channel Modeling}
Since the wavelength of THz waves is much shorter than the size of typical interacting objects, the dominant propagation mechanism is specular reflection, which can be effectively predicted using deterministic prediction methods like RT. However, the shape and surface roughness of interacting objects cannot be ignored compared to the wavelength. Therefore, a QD model appears to be more effective. 

For example, the impulse response of the QD model for a corridor environment developed in \cite{TAP} is expressed as 
\begin{eqnarray}
h_c(\tau, \phi_\T, \phi_\R) =   h_c^\d(\tau, \phi_\T, \phi_\R) + h_c^\r(\tau, \phi_\T, \phi_\R),
\end{eqnarray}
where $\tau$, $\phi_{\T}$, and $\phi_{\R}$ represent the delay time and angles of departure and arrival of an MPC, respectively. The direct path and specular reflection by walls are represented as deterministic components in $h_c^{\mathrm{d}}(\tau, \phi_\T, \phi_\R)$. On the other hand, the random components are expressed as 
\begin{eqnarray}
h_c^{\mathrm{r}}(\tau, \phi_\T, \phi_\R) = \sum \gamma_l^\r \delta (\tau_l^\r, \phi_{\T,l}^\r, \phi_{\R,l}^\r) \label{eq:rand}
\end{eqnarray}
where $\gamma_l^\r$ denotes the complex amplitude, and $\delta(\tau_l, \phi_{\T,l}, \phi_{\R,l}) \equiv \delta(\tau - \tau_l) \delta(\phi_\T - \phi_{\T, l}) \delta(\phi_\R - \phi_{\R, l})$. 

Moreover, as described above, diffuse scattering caused by surface irregularities and roughness can be modeled as a cluster of MPCs whose delays and angles are closely aligned with those of the associated specular component. By treating surface roughness as a spatially stationary stochastic process, stochastic modeling can effectively capture the small-scale fading of the specular reflection component resulting from interference with the diffuse scattering.

\subsection{Leveraging Multipaths for Multi-Stream Transmission}
The results presented above reveal that the richness and strength of multipath clusters are highly sensitive to the physical environment; specifically, larger and more open areas tend to yield fewer and weaker multipath components. In~\cite{WCL}, the feasibility of spatial multiplexing at the THz band is investigated in an open square scenario. It is observed that, on average, approximately $3.5$ multipath clusters with non-negligible power are present, indicating the potential for MIMO communication. Based on the measurement results, the benefit of utilizing multipath channels is quantified in terms of the achievable average channel capacity, with and without PRS. Here, the PRS made of metal is assumed to be present at the interacting points of the multipaths (clusters) to compensate for the interaction loss. The study shows that, even in the absence of PRS, an average channel capacity exceeding 9~bps/Hz in a $4 \times 4$ multi-beam MIMO configuration can be achieved in LoS scenarios, provided a 20~dB SNR is available at an approximately 10~m separation distance. Furthermore, with PRS, the average channel capacity reaches up to 18~bps/Hz. An accurate multipath cluster model is essential for proper link- and system-level evaluations, which are critical to the realization of robust and reliable THz communication networks.


\section{Human Body Shadowing Measurement and Modeling}
The propagation of THz signals is significantly affected by obstructions, including moving people, building walls, and furniture. In particular, a thorough investigation of the relationship between human motion and dynamic channel behavior is essential. This section introduces an in-house integrated measurement system developed for the simultaneous acquisition of channel responses and human motion, and proposes an RT-based HBS prediction model. The proposed model enables accurate modeling of diffraction and creeping effects, thereby enhancing the prediction accuracy of Doppler signatures resulting from human-induced blockage in RT simulations. Experimental validation of the proposed approach is also provided. Accurate prediction of HBS channels facilitates proactive path scheduling and improves link reliability, thereby supporting the development of robust THz communication systems.

\subsection{Motion Capture–Integrated Channel Sounder}
An integrated system for simultaneous acquisition of THz channel responses and human motion has been developed \cite{KangHBS300}. Detailed human motion is captured as point clouds using a motion capture (MoCap) system. The MoCap system and the channel sounder are synchronized via an external trigger. A continuous wave (CW) signal at $300$~GHz was employed for measurement. The received signal, influenced by human motion, was downconverted to $6$~GHz and sampled by the signal analyzer at a rate of $30$~kHz to capture the small-scale fading fluctuation caused by HBS. Multipath components from the surrounding environment were minimized by using $26$~dBi rectangular horn antennas on both the Tx and Rx ends. Due to limited output power, the EIRP was approximately $16$~dBm, providing a dynamic range of $54$~dB with an antenna separation of $3.5$~m.

\subsection{Accurate RT-based Model for HBS Propagation Channel}
Prediction techniques for scattering problems are widely used in modeling HBS propagation channels. Among these, full-wave approaches (i.e., EM-based approaches), which directly solve Maxwell’s equations, can produce highly accurate solutions, closely matching the exact behavior of electromagnetic waves. However, their high computational cost becomes a major limitation for large-scale problems or real-time computation, particularly at mm-wave and THz frequencies. As a cost-effective alternative, RT-based models are often preferred. The Uniform Geometrical Theory of Diffraction (UTD), which approximates individual electromagnetic propagation mechanisms such as edge diffraction, creeping diffraction, and reflection, is widely used in RT-based simulations. To overcome the limitations of treating these mechanisms separately, a unified formulation was proposed, integrating multiple propagation effects into a single analytical framework \cite{Du}. In this formulation, reflection or creeping diffraction from a curved object is expressed as the sum of an equivalent edge diffraction term and an additional correction term. This unified model has been validated against the exact solution for scattering from a dielectric circular cylinder at high frequencies. Importantly, this formulation not only unifies diffraction and reflection analytically but also enables the investigation of conditions under which the human body can be effectively modeled as a screen in high-frequency electromagnetic propagation scenarios.

\begin{figure}[t]
\centering
\includegraphics[width=0.99\linewidth]{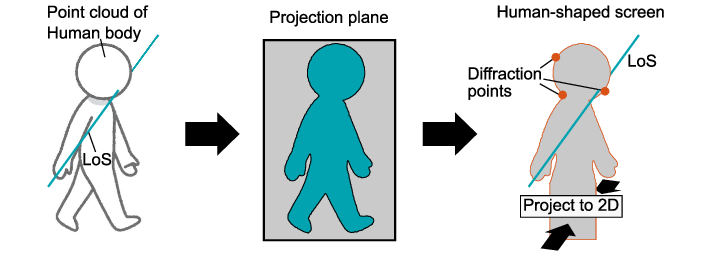} 
\caption{Human-shaped screen model. \label{fig:processes_HBscreen}}
\end{figure}

Since human blockage significantly affects the performance of LoS links at THz frequencies, switching to alternative propagation paths is a promising approach to mitigate the resulting shadowing effects. To ensure stable mobile communication, it is important to predict the shadowing event in advance, specifically within the lit region, in order to determine the optimal timing for path switching. In the lit region, the study in \cite{Du} showed that the proposed additional term can be neglected at high frequencies, and hence reflection from a dielectric cylinder can be effectively approximated as edge diffraction. Consequently, the dynamic HBS channel can be modeled as diffraction paths from a human, suitable for RT-based simulation. In \cite{KangHBS300}, a human-shaped screen model as illustrated in Fig.~\ref{fig:processes_HBscreen} was proposed to enhance prediction accuracy. This human-shaped screen is created by projecting the human body’s point cloud onto a projection plane which is perpendicular to the LoS. 
Since the phase of the diffraction paths is stationary, the criterion for identifying stationary phase points \cite{EYEfunc} can be applied to determine the diffraction points. 

\begin{figure*}[t]
\centering
\subfigure[Evaluation methodology. \label{fig:env_Office}]{\includegraphics[width=0.5\linewidth]{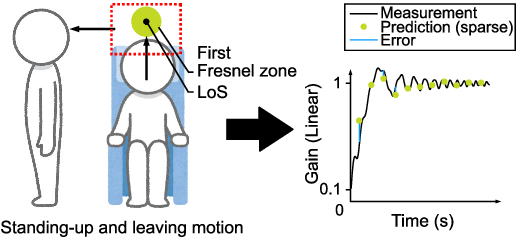}} 
\subfigure[Prediction error of the envelope of the fading pattern in the lit region. The diffraction paths from the human-shaped screen reduce the bias error by a factor of five. \label{fig:HM_Lit}]{\includegraphics[width=0.32\linewidth]{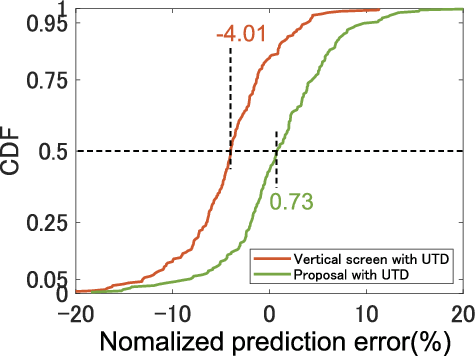}} 
\caption{Validation in an office scenario.\label{figs:HBS2}}
\end{figure*} 

\subsection{Validation in an Office Scenario}

The proposed HBS channel, modeled using the diffraction paths from the human-shaped screen as described above, was validated by comparing the channel gain fluctuations caused by the Doppler effect in the lit region between the predicted results and the measurements. Here, the Doppler effect arises from movement-induced changes in the phase and frequency of diffracted waves. The evaluation methodology is illustrated in Fig.~\ref{fig:env_Office}. In the measurement setup, the human subject was seated between the Tx and Rx antennas, representing one of the typical motion scenarios commonly observed in office environments. The proposed HBS channel model was compared with a conventional model based on a rectangular screen, where the prediction error, normalized by the LoS envelope, was evaluated. Due to the limited frame rate of the MoCap system, data from ten distinct measurements were collected for statistical comparison. The resulting cumulative distribution is shown in Fig.~\ref{fig:HM_Lit}. The results indicate that the proposed HBS channel model, using diffraction waves derived from a human-shaped screen, offers greater accuracy in terms of bias error compared to the rectangular screen model. Therefore, RT-based simulation, when employing the proposed HBS channel model, is demonstrated to be a reliable and accurate alternative to the EM-based approach. This enables light detection and ranging (LiDAR)-incorporated base stations to predict link context in real time by utilizing environmental point cloud data, thereby facilitating reliable transmission in 6G THz networks.

\section{Future Research Challenges}
The exploration of dynamic channel properties in indoor and outdoor environments, crucial for short-range mobile access in future communication systems, is necessary to understand the potential performance of the THz systems, but remains limited due to the absence of extensive measurement campaigns. In addition, the current measurement campaigns have primarily focused on frequencies below 500~GHz, leaving large portions of the spectrum uninvestigated. Furthermore, the dynamic behavior of channels, such as human body blocking, reflections, and scattering from mobile objects, needs further study. 

As discussed earlier, integrating IRS into THz communication systems is essential, offering a promising low-cost approach to addressing the short-range limitations and achieving reliable wireless connectivity. Moreover, IRS-assisted multi-beamforming can significantly enhance spectrum and energy efficiency while providing alternative propagation paths to mitigate the impact of human blockages. Importantly, due to extremely small wavelengths, the THz communications in short-range mobile scenarios will often have to operate in the near-field. In this scenario, the IRS can focus energy precisely on a specific mobile user, significantly enhancing communication efficiency. However, achieving this requires accurate near-field channel estimation and precise phase control of IRS unit cells, which remain critical challenges to be addressed. As THz communication matures, integrating advanced technologies like IRS, holographic MIMO, and heterogeneous 3D networks into dynamic short-distance scenarios will necessitate comprehensive studies to optimize frequency planning, communication range, and network capacity.

\section{Concluding Remarks}
While THz frequency bands offer the potential for ultra-high-speed data transmission at rates of hundreds of gigabits per second, enabled by their large channel bandwidth, concrete applications and usage scenarios have yet to be fully articulated. This is mainly due to the inherent challenges associated with THz frequencies, including limited coverage, significant path losses, and human blockages. This article has addressed key issues spanning static multipath channel behavior and dynamic blockage modeling, contributing integrated insights into both aspects of THz channel characterization. Moving forward, an extensive system-level evaluation that unifies these perspectives will be essential to realizing robust and reliable THz communication networks.


\IEEEbiographynophoto{Minseok Kim}
is a Principal Investigator and Research Professor at Niigata University, Japan. His current research focuses on 6G mmWave/THz communications. He serves as the Chair of the IEEE ComSoc P1944 THz Subgroup and as an Associate Editor for \textit{IEEE Antennas Wirel. Propag. Lett.}
\vspace{-3mm}

\IEEEbiographynophoto{Jun-ichi Takada}
serves as the Executive Vice President for Global Affairs and is a professor at the School of Environment at the Institute of Science Tokyo (Science Tokyo), Tokyo, Japan Society.
\vspace{-3mm}

\IEEEbiographynophoto{Minghe Mao}
is a specially appointed Assistant Professor at Niigata University, Japan. His current focus is on 6G THz communications.
\vspace{-3mm}

\IEEEbiographynophoto{Che Chia Kang}
received the D.E. degree from the Tokyo Institute of Technology, Tokyo, Japan, in 2024. His research interests include numerical electromagnetic simulation and measurements of the human shadowing effect.
\vspace{-3mm}

\IEEEbiographynophoto{Xin Du}
is an Assistant Professor at Kagoshima University, Japan. His research interests include numerical electromagnetic simulation, diffraction theory, and reconfigurable intelligent surfaces. 
\vspace{-3mm}

\IEEEbiographynophoto{Anirban Ghosh}
is an Assistant Professor from SRM University AP, Andhra Pradesh, India.
\vspace{-3mm}

\end{document}